\documentclass[aip,apl,reprint,amsmath,amssymb]{revtex4-1}

\usepackage{graphicx}
\usepackage{bm}
\usepackage[utf8]{inputenc}
\usepackage[T1]{fontenc}
\usepackage{mathptmx}
\usepackage{etoolbox}
\usepackage[svgnames]{xcolor}
\usepackage{hyperref}
\hypersetup{colorlinks,citecolor=DarkGreen,linkcolor=FireBrick,urlcolor=FireBrick,linktocpage,unicode}
\usepackage[normalem]{ulem}
\usepackage{cancel}

\usepackage{braket}
\allowdisplaybreaks[3]
\usepackage[mathlines]{lineno}

\newcommand{\TA}[1]{\textcolor{Black}{#1}}
\DeclareRobustCommand{\TAdel}{\bgroup\markoverwith{\textcolor{Green}{\rule[.5ex]{2pt}{1pt}}}\ULon}

\newcounter{hoge}

\makeatletter
\@addtoreset{equation}{hoge}
\makeatother

\begin{document}


\title{Residual-\texorpdfstring{$ZZ$}{ZZ}-coupling suppression and fast two-qubit gate for Kerr-cat qubits based on level-degeneracy engineering}

\author{Takaaki Aoki}
\email{takaaki-aoki@aist.go.jp}
\affiliation{Global Research and Development Center for Business by Quantum-AI Technology (G-QuAT), National Institute of Advanced Industrial Science and Technology (AIST), 1-1-1 Umezono, Tsukuba, Ibaraki 305-8568, Japan}

\author{Akiyoshi Tomonaga}
\affiliation{Global Research and Development Center for Business by Quantum-AI Technology (G-QuAT), National Institute of Advanced Industrial Science and Technology (AIST), 1-1-1 Umezono, Tsukuba, Ibaraki 305-8568, Japan}
\affiliation{NEC-AIST Quantum Technology Cooperative Research Laboratory, National Institute of Advanced Industrial Science and Technology (AIST), 1-1-1 Umezono, Tsukuba, Ibaraki 305-8568, Japan}

\author{Kosuke Mizuno}
\affiliation{Global Research and Development Center for Business by Quantum-AI Technology (G-QuAT), National Institute of Advanced Industrial Science and Technology (AIST), 1-1-1 Umezono, Tsukuba, Ibaraki 305-8568, Japan}

\author{Shumpei Masuda}
\email{shumpei.masuda@aist.go.jp}
\affiliation{Global Research and Development Center for Business by Quantum-AI Technology (G-QuAT), National Institute of Advanced Industrial Science and Technology (AIST), 1-1-1 Umezono, Tsukuba, Ibaraki 305-8568, Japan}
\affiliation{NEC-AIST Quantum Technology Cooperative Research Laboratory, National Institute of Advanced Industrial Science and Technology (AIST), 1-1-1 Umezono, Tsukuba, Ibaraki 305-8568, Japan}
\date{\today}

\begin{abstract}
    Building large-scale quantum computers requires an interqubit-coupling scheme with a high on-off ratio to avoid unwanted crosstalk coming from residual coupling and to enable fast multi-qubit operations.
    We propose a $ZZ$-coupling scheme for two Kerr-cat qubits with a frequency-tunable  coupler.
    By making four relevant states of the two Kerr-cat qubits quadruply degenerate, we can switch off the $ZZ$ coupling.
    By partially lifting the level degeneracy, we can switch it on.
    We theoretically show that an experimentally feasible circuit model suppresses the residual $ZZ$ coupling. 
    Moreover, our circuit can realize $R_{ZZ}(-\pi/2)$-gate fidelity higher than $99.9\%$ within $\TA{18}$\,ns when decoherence is ignored.
    \TA{Our model includes the first-order terms in expansion beyond the rotating-wave approximation.}
\end{abstract}


\maketitle 

A Kerr-cat qubit, which stores quantum information on a \TA{parametrically} \TA{(}squeeze-\TA{)}driven Kerr-nonlinear oscillator \TA{(KPO)},\cite{Cochrane1999,GotoPRA2016,Puri2017} is attracting much attention as a candidate platform for quantum computation.
One of its advantages is its biased-noise nature\TA{: its logical states are defined as two coherent states with opposite phases, and the bit-flip rate is exponentially suppressed with the mean photon number},\cite{PhysRevX.9.041009} which allows efficient quantum error corrections.\cite{Darmawan2021}
\TA{This biased-noise nature comes from a double-well Hamiltonian of a KPO.\cite{PhysRevX.14.031040}
To investigate its energy-level structure, the reflection spectroscopy has been studied theoretically\cite{Masuda_2021_9_15} and experimentally.\cite{Yamaguchi_2024}
Energies of an effective static Hamiltonian of a KPO and quasi-energies obtained from Floquet theory have been compared theoretically.\cite{GarciaMata2024effectiveversus}
Pairwise level degeneracies owing to increased barrier height of the double well have been observed in experiment.\cite{PhysRevX.14.031040,Chavez-Carlos2023}
When the detuning of the resonance frequency of a KPO from half the parametric-drive frequency takes specific values, the energy spectrum of the KPO shows multiple degeneracies and bit-flip errors are further suppressed, which has been shown theoretically\cite{PhysRevA.107.042407} and experimentally.\cite{doi:10.1073/pnas.2311241121,2024arXiv241104442Q}
A fully tunable asymmetric double well has been created experimentally.\cite{2024arXiv240913113C}}

A universal gate set for Kerr-cat qubits can be constructed for example by Z-axis rotations ($R_Z$ gates) with all rotation angles, an X-axis rotation ($R_X$ gate) with rotation angle $-\pi/2$, and a ZZ rotation ($R_{ZZ}$ gate) with rotation angle $-\pi/2$.\cite{GotoPRA2016}
Various schemes for gate operations on Kerr-cat qubits have been \TA{theoretically} studied:
an $R_Z$ gate with a single-photon drive;\cite{GotoPRA2016,Puri2017}
an $R_X$ gate by controlling the oscillator frequency;\cite{GotoPRA2016,Puri2017}
an $R_X$ gate using time evolution under the Kerr Hamiltonian without the squeezing drive;\cite{Puri2017}
an $R_X$ gate using effective excited states;\cite{PhysRevApplied.18.014019}
an $R_{ZZ}$ gate using beam-splitter coupling;\cite{GotoPRA2016,Puri2017}
an $R_{ZZ}$ gate by controlling the phase of the squeezing drive;\cite{MasudaPRApplied2022}
an $R_{ZZ}$ gate using conditional driving;\cite{Chono2022,2024arXiv241000552C}
acceleration of the elementary gates;\cite{PhysRevResearch.6.013192}
\TA{and} nontrivial bias-preserving gates.\cite{PuriSA2020,PhysRevApplied.18.024076,PhysRevResearch.4.013082_2022_2_2}
\TA{Gate operations on Kerr-cat qubits have also been experimentally demonstrated:
$R_Z$, $R_X$,\cite{Iyama2024} and $\sqrt{\mathrm{iSWAP}}$\cite{2024arXiv240617999H} gates using dc superconducting quantum interference devices (SQUIDs); and
$R_Z$ and $R_X$ gates using superconducting nonlinear asymmetric inductive elements (SNAILs).\cite{Grimm,PhysRevX.14.041049,2024arXiv241104442Q}}

An $R_{ZZ}$ gate is based on $ZZ$ coupling between qubits.
If we cannot switch off the $ZZ$ coupling, the residual coupling causes crosstalk\cite{Sarovar2020detectingcrosstalk} among qubits, yielding unwanted correlations between them.
Because crosstalk errors are difficult to remove with quantum error corrections, which in general depend on errors to be local,\cite{Sarovar2020detectingcrosstalk} a scheme without residual coupling is desired to build a large-scale quantum computer.
In order to suppress residual coupling,  tunable couplers have been utilized in transmon systems,\cite{PhysRevLett.113.220502,PhysRevApplied.10.054062,PhysRevApplied.12.054023,PhysRevApplied.14.024070,PhysRevLett.125.200503,PhysRevLett.125.200504,PhysRevLett.127.080505,PhysRevApplied.16.064062,PhysRevApplied.18.034038,PRXQuantum.3.020301,PRXQuantum.4.010314,10.1063/5.0138699,PhysRevApplied.22.024057} which also enable fast multi-qubit operations. 
In Ref.~\onlinecite{Aoki2024}, the authors have developed a coupling scheme for two Kerr-cat qubits in which two tunable resonators are used as couplers and the frequency of a resonator is controlled.
Although the scheme can suppress the residual coupling, a simpler scheme with less residual coupling is desirable.

In this Letter we propose a coupling scheme for two Kerr-cat qubits with a better performance using a single tunable resonator as a coupler.
By tuning the coupler frequency, we engineer the degeneracy of four relevant states of the two Kerr-cat qubits.
This allows a cancellation of the residual $ZZ$ coupling and a fast and high-fidelity $R_{ZZ}$ gate.
We show a circuit model that is experimentally feasible and numerically investigate its performance.
\TA{Our Hamiltonian derived from the circuit model incorpotrates the first-order terms in expansion beyond the rotating-wave approximation (RWA).}
\TA{We use two coherent states with opposite phases as the computational basis of a Kerr-cat qubit in this Letter, although their superpositions are also used as computational basis in other papers.}

Before discussing the scheme for Kerr-cat qubits, we explain the relation between $ZZ$ coupling and energy-level degeneracy of a diagonalized two-qubit Hamiltonian. 
The Hamiltonian is given by
\begin{linenomath}
\begin{align}
    \hat{H}_{2\mathrm{q}}=\sum_{l,m=0}^1E_{l,m}\Ket{\psi_{l,m}}\Bra{\psi_{l,m}}=\begin{pmatrix}
        E_{0,0} & 0 & 0 & 0 \\
        0 & E_{0,1} & 0 & 0 \\
        0 & 0 & E_{1,0} & 0 \\
        0 & 0 & 0 & E_{1,1} \\
    \end{pmatrix},
    \label{eq:H2q1}
\end{align}
\end{linenomath}
where $\left.\left\{\Ket{\psi_{l,m}}\right|l,m\in\{0,1\}\right\}$ are four eigenstates and form an orthonormal basis; $E_{l,m}$ is the eigenenergy of eigenstate $\Ket{\psi_{l,m}}$.
This Hamiltonian can be rewritten as
\begin{linenomath}
\begin{align}
    \hat{H}_{2\mathrm{q}}/\hbar=\frac{\zeta_{II}}{4}\hat{I}\hat{I}+\frac{\zeta_{ZZ}}{4}\hat{Z}\hat{Z}
    +\frac{\zeta_{ZI}}{4}\hat{Z}\hat{I}
    +\frac{\zeta_{IZ}}{4}\hat{I}\hat{Z},
\end{align}
\end{linenomath}
where
\begin{linenomath}
\begin{align}
    \hbar\zeta_{II}=E_{0,0}+E_{0,1}+E_{1,0}+E_{1,1}, \label{eq:zetaII}\\
    \hbar\zeta_{ZZ}=E_{0,0}-E_{0,1}-E_{1,0}+E_{1,1}, \label{eq:zetaZZ} \\
    \hbar\zeta_{ZI}=E_{0,0}+E_{0,1}-E_{1,0}-E_{1,1}, \label{eq:zetaZI} \\
    \hbar\zeta_{IZ}=E_{0,0}-E_{0,1}+E_{1,0}-E_{1,1}, \label{eq:zetaIZ} \\
    \hat{I}=\begin{pmatrix}
        1 & 0 \\
        0 & 1 \\
    \end{pmatrix}, \quad
    \hat{Z}=\begin{pmatrix}
        1 & 0 \\
        0 & -1 \\
    \end{pmatrix},
\end{align}
\end{linenomath}
and $\hbar=h/(2\pi)$ is the reduced Planck constant.
$\zeta_{ZZ}$ in Eq.~\eqref{eq:zetaZZ} is a $ZZ$-coupling strength.\cite{PhysRevApplied.12.054023,PhysRevApplied.14.024070,PhysRevLett.125.200503,PhysRevLett.125.200504,PhysRevLett.127.080505,PhysRevApplied.16.064062,PhysRevApplied.18.034038,PRXQuantum.3.020301,PRXQuantum.4.010314,10.1063/5.0138699,PhysRevApplied.22.024057}
Our strategy to cancel $ZZ$ coupling, $\zeta_{ZZ}=0$, is to make the four eigenstates quadruply degenerate by tuning system parameters, which also leads to $\zeta_{ZI}=0$ and $\zeta_{IZ}=0$.
We define four logical states $\left\{\left.\Ket{\widetilde{l,m}}\right|l,m\in\{0,1\}\right\}$ as the quadruply degenerate eigenstates.

\begin{figure}
    \centering
    \includegraphics[width=0.45\textwidth]{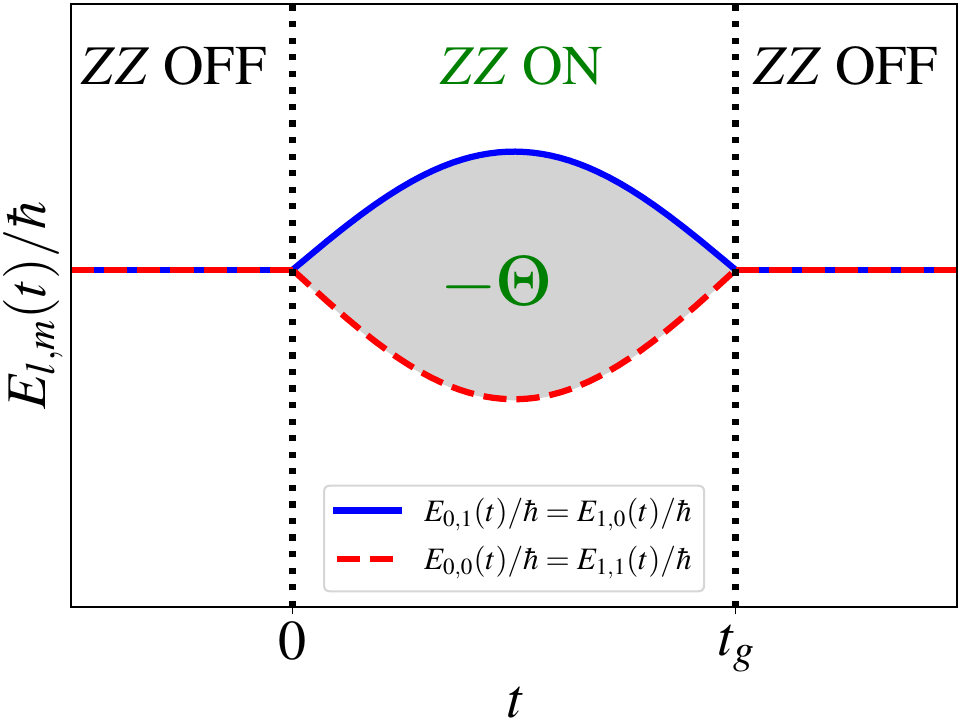}
    \caption{A schematic of our level-degeneracy engineering to control $ZZ$ coupling.
    An $R_{ZZ}(\Theta)$ gate is applied for $0\leq t\leq t_g$.
    The light gray area is $-\Theta$.
    \label{fig:ZZschematic}}
\end{figure}

On the other hand, if we retain the degeneracy between $\Ket{\psi_{0,0}}$ and $\Ket{\psi_{1,1}}$ and that between $\Ket{\psi_{0,1}}$ and $\Ket{\psi_{1,0}}$ while partially lifting the degeneracy between the former two states and the latter two, we can perform only an $R_{ZZ}$ gate as follows.
We assume that at $t=0$, $ZZ$ coupling is switched off, $\zeta_{ZZ}(0)=0$.
We prepare the initial state as
\begin{linenomath}
\begin{align}
    \Ket{\Psi(0)}=\sum_{l,m=0}^1\beta_{l,m}\Ket{\widetilde{l,m}},
    \label{eq:initial1}
\end{align}
\end{linenomath}
where $\beta_{l,m}$ is a coefficient.
When the system parameters are changed adiabatically with the condition $\zeta_{ZZ}(t_g)=0$, where $t_g$ is the gate time, the state of the system at $t=t_g$ becomes
\begin{linenomath}
\begin{align}
    \Ket{\Psi(t_g)}&=\mathcal{T}\exp\left(
        -\frac{\mathrm{i}}{\hbar}\int_0^{t_g}\hat{H}_{2\mathrm{q}}(t)\,\mathrm{d}t
    \right)\Ket{\Psi(0)}
    \notag \\
    &=\mathrm{e}^{-\mathrm{i}\theta}\hat{R}_{ZZ}(\Theta)\ket{\Psi(0)}
    =:\mathrm{e}^{-\mathrm{i}\theta}\Ket{\Psi_{\Theta}^{\mathrm{ideal}}},
    \label{eq:state_tg}
\end{align}
\end{linenomath}
where $\mathcal{T}$ is the time-ordering operator,
\begin{linenomath}
\begin{align}
    \theta:=\int_0^{t_g}\frac{\zeta_{II}(t)}{4}\,\mathrm{d}t
\end{align}
\end{linenomath}
is a global phase, and
\begin{linenomath}
\begin{align}
    \hat{R}_{ZZ}(\Theta)&=\sum_{l,m=0}^1\mathrm{e}^{-\mathrm{i}(2\delta_{l,m}-1)\Theta/2}\Ket{\widetilde{l,m}}\Bra{\widetilde{l,m}}
    \notag \\
    &=\begin{pmatrix}
        \mathrm{e}^{-\mathrm{i}\Theta/2} & 0 & 0 & 0 \\
        0 & \mathrm{e}^{\mathrm{i}\Theta/2} & 0 & 0 \\
        0 & 0 & \mathrm{e}^{\mathrm{i}\Theta/2} & 0 \\
        0 & 0 & 0 & \mathrm{e}^{-\mathrm{i}\Theta/2} \\
    \end{pmatrix}
\end{align}
\end{linenomath}
with
\begin{linenomath}
\begin{align}
    \Theta:=\int_0^{t_g}\frac{\zeta_{ZZ}(t)}{2}\,\mathrm{d}t
\end{align}
\end{linenomath}
being the rotation angle and $\delta_{l.m}$ being the Kronecker delta.
Here, we have ignored decoherence.
In reality, some unwanted nonadiabatic transitions are unavoidable, and the second equal sign in Eq.~\eqref{eq:state_tg} is replaced by an approximately equal one.
For evaluation of the degree of approximation, we later calculate  gate fidelity.
We show a schematic of our level-degeneracy engineering in Fig.~\ref{fig:ZZschematic}.

\begin{figure}
    \centering
    \includegraphics[width=0.38\textwidth]{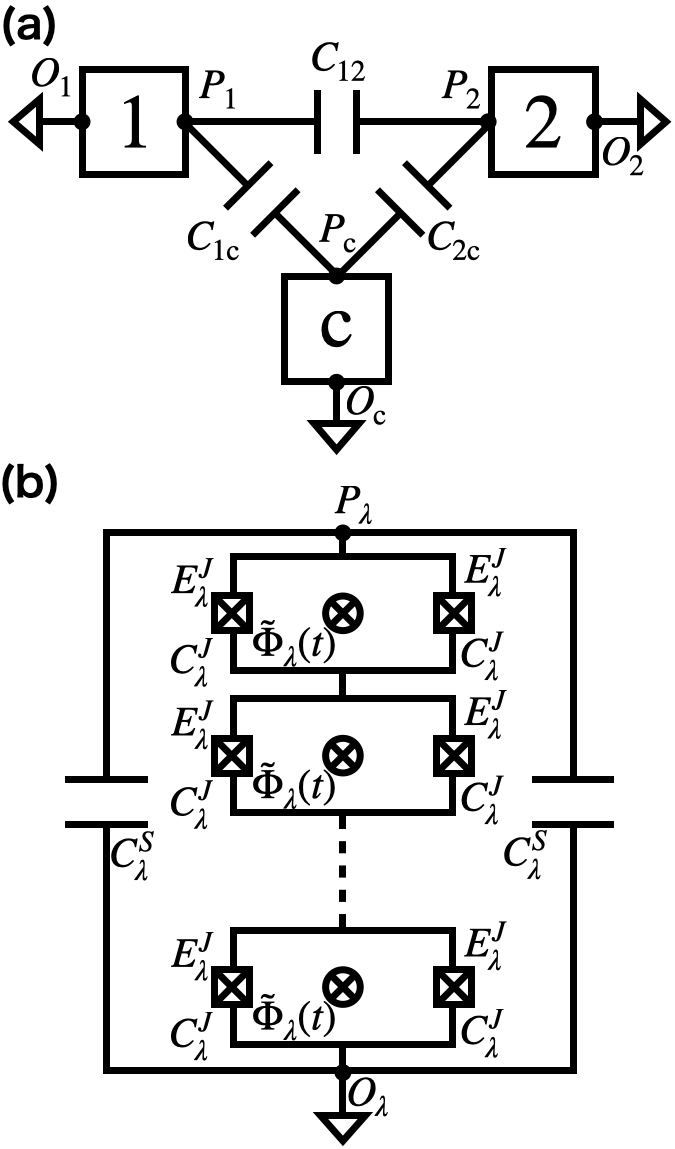}
    \caption{A circuit model of our system, which constitutes two Kerr-cat qubits.
    (a) Subsystems $1$, $2$, and c are coupled capacitively.
    (b) A circuit diagram of subsystem $\lambda\in\{1,2,\mathrm{c}\}$, which comprises two identical shunting capacitors of capacitance $C_{\lambda}^S$ and an array of $N_{\lambda}$ identical symmetric dc SQUIDs each of which has two Josephson junctions of Josephson energy $E_{\lambda}^J$ and capacitance $C_{\lambda}^J$.
    It is connected to the ground via node $O_{\lambda}$.
    It is symmetric with respect to the line through $P_{\lambda}$ and $O_{\lambda}$.
    The center of each SQUID is threaded by a magnetic flux $\tilde{\Phi}_{\lambda}(t)$ into the paper.\cite{2022arXiv220101945N}
     \label{fig:system1}}
\end{figure}
Now, we introduce the level-degeneracy engineering for Kerr-cat qubits.
We consider a system consisting of two Kerr parametric oscillators~\cite{GotoPRA2016,Puri2017} (KPOs, subsystems $1$ and $2$) and a tunable resonator (subsystem c).
This system constitutes two Kerr-cat qubits.
A circuit model of our system is shown in Fig.~\ref{fig:system1}.
The center of each SQUID of subsystem $\lambda\in\{1,2,\mathrm{c}\}$ is threaded by a magnetic flux $\tilde{\Phi}_{\lambda}(t)$.
We decompose a dimensionless magnetic flux $\tilde{\varphi}_{\lambda}(t):=\tilde{\Phi}_{\lambda}(t)/\phi_0$, where $\phi_0=\hbar/(2e)$ is the reduced flux quantum [$h/(2e)$ is the flux quantum], into bias and pump parts as\cite{GotoJPSJ2019}
\begin{linenomath}
\begin{gather}
    \tilde{\varphi}_{\lambda}(t)=\tilde{\varphi}_{\lambda}^{\mathrm{bias}}(t)+\tilde{\varphi}_{\lambda}^{\mathrm{pump}}(t), \\
    \tilde{\varphi}_{\lambda}^{\mathrm{pump}}(t)
    =-2\pi\epsilon_{p,\lambda}\cos{(\omega_pt)},
\end{gather}
\end{linenomath}
with $\epsilon_{p,j}\ll1$ for $j\in\{1,2\}$ and $\epsilon_{p,\mathrm{c}}=0$.
In the laboratory frame, the KPOs are parametrically pumped at frequency $\omega_p$ but the resonator is not.
The bias parts $\{\tilde{\varphi}_{\lambda}^{\mathrm{bias}}(t)\}$ are the key components for the level-degeneracy engineering.
At first, we assume that they are time independent; $\tilde{\varphi}_{\lambda}^{\mathrm{bias}}(t)=\tilde{\varphi}_{\lambda}^{\mathrm{bias}}(0)=\tilde{\varphi}_{\lambda}^{\mathrm{bias}}\;\forall t$. 
The \TA{effective} Hamiltonian $\hat{H}$ of our system is written as (see the supplementary material)
\begin{linenomath}
\begin{align}
    \hat{H}&=\sum_{j=1,2}\hat{H}_{j}+\hat{H}_{\mathrm{c}}+\hat{H}_{\mathrm{I}}, \label{eq:HF2} \\
    \hat{H}_{j}/\hbar&=-\frac{\tilde{K}_j}{2}\hat{a}_j^{\dag2}\hat{a}_j^2+\frac{\tilde{p}_j}{2}(\hat{a}_j^{\dag2}+\hat{a}_j^2)+\tilde{\Delta}_j\hat{a}_j^{\dag}\hat{a}_j
    \label{eq:HF2KPO}
    \notag \\
    &\quad\mbox{}-\frac{17K_j^2}{18\omega_p}\hat{a}_j^{\dag3}\hat{a}_j^3+\frac{5K_jp_j}{6\omega_p}(\hat{a}_j^{\dag3}\hat{a}_j+\hat{a}_j^{\dag}\hat{a}_j^3), \\
    \hat{H}_{\mathrm{c}}/\hbar&=-\frac{\tilde{K}_{\mathrm{c}}}{2}\hat{a}_{\mathrm{c}}^{\dag2}\hat{a}_{\mathrm{c}}^2+\tilde{\Delta}_{\mathrm{c}}\hat{a}_{\mathrm{c}}^{\dag}\hat{a}_{\mathrm{c}}-\frac{17K_{\mathrm{c}}^2}{18\omega_p}\hat{a}_{\mathrm{c}}^{\dag3}\hat{a}_{\mathrm{c}}^3, \\
    \hat{H}_{\mathrm{I}}/\hbar&=\sum_{j=1,2}\tilde{g}_{j\mathrm{c}}(\hat{a}_j^{\dag}\hat{a}_{\mathrm{c}}+\hat{a}_j\hat{a}_{\mathrm{c}}^{\dag})
    +\tilde{g}_{12}(\hat{a}_1^{\dag}\hat{a}_2+\hat{a}_1\hat{a}_2^{\dag}) \notag \\
    &\quad\mbox{}-\sum_{j=1,2}\frac{g_{j\mathrm{c}}}{\omega_p}[K_j(\hat{a}_j^{\dag2}\hat{a}_j\hat{a}_{\mathrm{c}}+\hat{a}_j^{\dag}\hat{a}_j^2\hat{a}_{\mathrm{c}}^{\dag})
    \notag \\
    &\quad\mbox{}-p_j(\hat{a}_j^{\dag}\hat{a}_{\mathrm{c}}^{\dag}+\hat{a}_j\hat{a}_{\mathrm{c}})+K_{\mathrm{c}}(\hat{a}_j^{\dag}\hat{a}_{\mathrm{c}}^{\dag}\hat{a}_{\mathrm{c}}^{2}+\hat{a}_j\hat{a}_{\mathrm{c}}^{\dag2}\hat{a}_{\mathrm{c}})] \notag \\
    &\quad\mbox{}-\frac{g_{12}}{\omega_p}[K_1(\hat{a}_1^{\dag2}\hat{a}_1\hat{a}_2+\hat{a}_1^{\dag}\hat{a}_1^2\hat{a}_2^{\dag})
    \notag \\
    &\quad\mbox{}+K_2(\hat{a}_1^{\dag}\hat{a}_2^{\dag}\hat{a}_2^2+\hat{a}_1\hat{a}_2^{\dag2}\hat{a}_2)-(p_1+p_2)(\hat{a}_1^{\dag}\hat{a}_2^{\dag}+\hat{a}_1\hat{a}_2)],
\end{align}
\end{linenomath}
where $\hat{H}_{\lambda}$ is the Hamiltonian of subsystem $\lambda\in\{1,2,\mathrm{c}\}$ and $\hat{H}_{\mathrm{I}}$ is the interaction Hamiltonian; $\hat{a}_{\lambda}$ is the annihilation operator of subsystem $\lambda\in\{1,2,\mathrm{c}\}$;
\TA{The parameters with tilde incorporate the effect of the first-order terms in expansion beyond the rotating-wave approximation; see the supplementary material;}
$K_{\lambda}$ is the Kerr nonlinearity of subsystem $\lambda\in\{1,2,\mathrm{c}\}$;
${p}_j$ is the amplitude of the parametric drive of subsystem $j\in\{1,2\}$;
$\Delta_{\lambda}$ is the detuning of the dressed resonance frequency $\omega_{\lambda}$ of subsystem $\lambda\in\{1,2,\mathrm{c}\}$ from $\omega_p/2$, that is, $\Delta_{\lambda}=\omega_{\lambda}-\omega_p/2$;
${g}_{\lambda\lambda'}$ is the coupling strength between subsystems $\lambda$ and $\lambda'$ ($\lambda\lambda'\in\{12,1\mathrm{c},2\mathrm{c}\}$).
The above parameters are tuned through $\{\tilde{\varphi}_{\lambda}^{\mathrm{bias}}\}$ as shown in the supplementary material.
\TA{We note that although $K_{\mathrm{c}}$ and $\tilde{K}_{\mathrm{c}}$ are not essential for our scheme, we include them in our Hamiltonian to make our model realistic, because a tunable resonator actually has Kerr nonlinearity; see, for example, Sec.~II.B.1 in Ref.~\onlinecite{doi:10.1063/1.5089550}.
In the literature on KPOs, as a Hamiltonian of a KPO, a simpler form, which is derived using the rotating-wave approximation and contains only the first line in Eq.~\eqref{eq:HF2KPO} without tilde, is often used.
However, when the Kerr nonlinearity is large, the discrepancy between the simple Hamiltonian and a Hamiltonian considering counter-rotating terms, which are omitted in the rotating-wave approximation, cannot be neglected.\cite{GarciaMata2024effectiveversus,Masuda2021,Chavez-Carlos_2025,2024arXiv240913113C}
This is why we have included the first-order terms in expansion beyond the rotating-wave approximation; see the supplementary material.}

\begin{figure}
    \centering
    \includegraphics[width=0.3\textwidth]{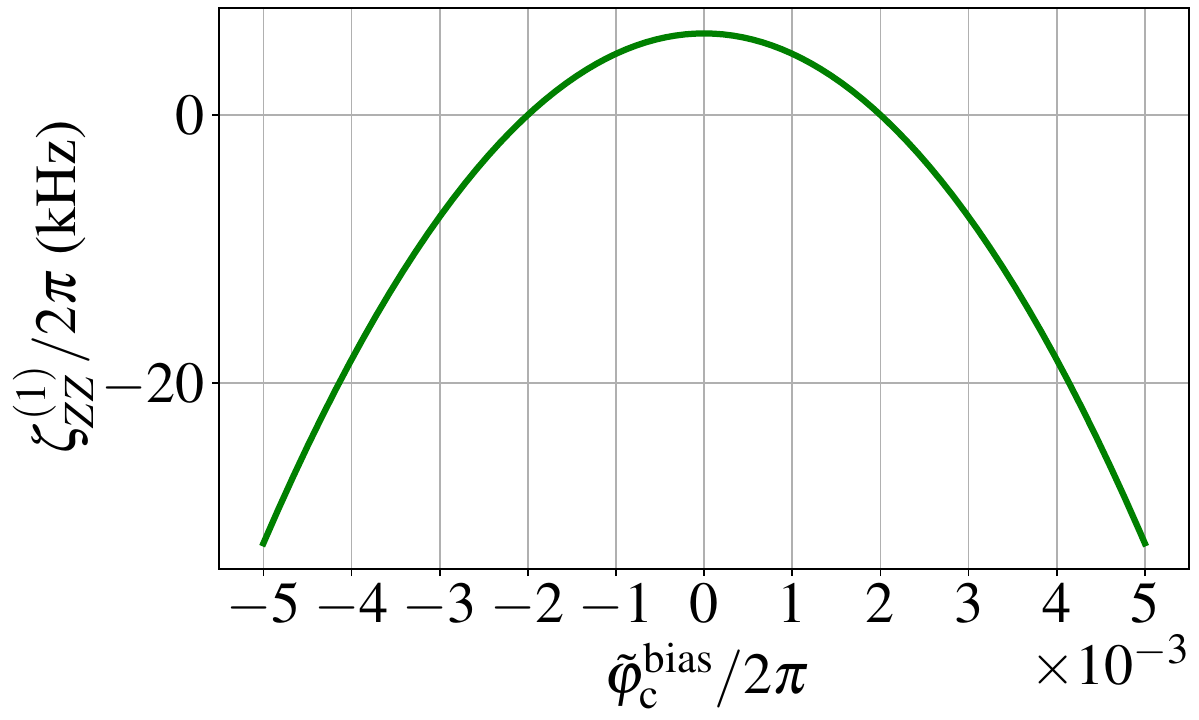}
    \caption{$\zeta^{\TA{(1)}}_{ZZ}$ as a function of $\tilde{\varphi}_{\mathrm{c}}^{\rm bias}$.
    We have $\zeta^{\TA{(1)}}_{ZZ}/2\pi=0$\,kHz when $\tilde{\varphi}_{\mathrm{c}}^{\rm bias}/2\pi\approx\pm2\times10^{-3}$.
    The relation between $\tilde{\varphi}_{\mathrm{c}}^{\rm bias}$ and system parameters are given in the supplementary material.
    Values of the parameters unrelated to $\tilde{\varphi}_{\mathrm{c}}^{\rm bias}$ are set as in Table \ref{table:parameter1}.
    \label{fig:zetaZZ}}
\end{figure}

Hamiltonian $\hat{H}$ can be rewritten as
\begin{align}
    \hat{H}&=\hat{H}_{0}+\hat{H}_{ZZ}+\sum_{j=1,2}\hat{H}_{X_j},\label{eq:HF3} \\
    \hat{H}_{0}/\hbar&=\sum_{j=1,2}\left[-\frac{\tilde{K}_j}{2}\left(\hat{a}_j^{\dag2}-\tilde{\alpha}_j^2\right)\left(\hat{a}_j^2-\tilde{\alpha}_j^2\right)+\frac{\tilde{K}_j}{2}\tilde{\alpha}_j^4\right]
    \notag \\
    &\quad\mbox{}+\tilde{\Delta}_{\mathrm{c}}\left(
        \hat{a}_{\mathrm{c}}^{\dag}+\frac{\tilde{g}_{1\mathrm{c}}}{\tilde{\Delta}_{\mathrm{c}}}\hat{a}_1^{\dag}+\frac{\tilde{g}_{2\mathrm{c}}}{\tilde{\Delta}_{\mathrm{c}}}\hat{a}_2^{\dag}
    \right)
    \notag \\
    &\quad\mbox{}\times
    \left(
        \hat{a}_{\mathrm{c}}+\frac{\tilde{g}_{1\mathrm{c}}}{\tilde{\Delta}_{\mathrm{c}}}\hat{a}_1+\frac{\tilde{g}_{2\mathrm{c}}}{\tilde{\Delta}_{\mathrm{c}}}\hat{a}_2
    \right), \label{eq:HF0} \\
    \hat{H}_{ZZ}/\hbar&=\left(\tilde{g}_{12}-\frac{\tilde{g}_{1\mathrm{c}}\tilde{g}_{2\mathrm{c}}}{\tilde{\Delta}_{\mathrm{c}}}\right)(\hat{a}_1^{\dag}\hat{a}_2+\hat{a}_1\hat{a}_2^{\dag})\notag \\
    &\quad\mbox{}-\frac{\tilde{K}_{\mathrm{c}}}{2}\hat{a}_{\mathrm{c}}^{\dag2}\hat{a}_{\mathrm{c}}^2
    -\frac{17K_{\mathrm{c}}^2}{18\omega_p}\hat{a}_{\mathrm{c}}^{\dag3}\hat{a}_{\mathrm{c}}^3
    \notag \\
    &\quad\mbox{}-\sum_{j=1,2}\frac{g_{j\mathrm{c}}}{\omega_p}[K_j(\hat{a}_j^{\dag2}\hat{a}_j\hat{a}_{\mathrm{c}}+\hat{a}_j^{\dag}\hat{a}_j^2\hat{a}_{\mathrm{c}}^{\dag})
    \notag \\
    &\quad\mbox{}-p_j(\hat{a}_j^{\dag}\hat{a}_{\mathrm{c}}^{\dag}+\hat{a}_j\hat{a}_{\mathrm{c}})+K_{\mathrm{c}}(\hat{a}_j^{\dag}\hat{a}_{\mathrm{c}}^{\dag}\hat{a}_{\mathrm{c}}^{2}+\hat{a}_j\hat{a}_{\mathrm{c}}^{\dag2}\hat{a}_{\mathrm{c}})] \notag \\
    &\quad\mbox{}-\frac{g_{12}}{\omega_p}[K_1(\hat{a}_1^{\dag2}\hat{a}_1\hat{a}_2+\hat{a}_1^{\dag}\hat{a}_1^2\hat{a}_2^{\dag})
    \notag \\
    &\quad\mbox{}+K_2(\hat{a}_1^{\dag}\hat{a}_2^{\dag}\hat{a}_2^2+\hat{a}_1\hat{a}_2^{\dag2}\hat{a}_2)
    \notag \\
    &\quad\mbox{}-(p_1+p_2)(\hat{a}_1^{\dag}\hat{a}_2^{\dag}+\hat{a}_1\hat{a}_2)],
    \label{eq:HFZZ} \\
    \hat{H}_{X_j}/\hbar&=\left(\tilde{\Delta}_j-\frac{\tilde{g}_{j\mathrm{c}}^2}{\tilde{\Delta}_{\mathrm{c}}}\right)\hat{a}_j^{\dag}\hat{a}_j-\frac{17K_j^2}{18\omega_p}\hat{a}_j^{\dag3}\hat{a}_j^3
    \notag \\
    &\quad\mbox{}+\frac{5K_jp_j}{6\omega_p}(\hat{a}_j^{\dag3}\hat{a}_j+\hat{a}_j^{\dag}\hat{a}_j^3), \label{eq:HFX}
\end{align}
where \TA{$\tilde{\alpha}_j:=\sqrt{\tilde{p}_j/\tilde{K}_j}$}, $\hat{H}_{ZZ}$ is a $ZZ$-coupling Hamiltonian, and $\hat{H}_{X_{\TA{j}}}$ is a Hamiltonian for $R_X$ gates \TA{on the $j$th Kerr-cat qubit}.
The following tensor products of coherent states of subsystems $1$, $2$, and c,
\begin{align}
    \Ket{\psi_{0,0}}&:=\Ket{\tilde{\alpha}_1,\tilde{\alpha}_2,-\tilde{\alpha}_{\mathrm{c}}^{+}}, \label{eq:eigenF00} \\
    \Ket{\psi_{0,1}}&:=\Ket{\tilde{\alpha}_1,-\tilde{\alpha}_2,-\tilde{\alpha}_{\mathrm{c}}^{-}}, \label{eq:eigenF01} \\
    \Ket{\psi_{1,0}}&:=\Ket{-\tilde{\alpha}_1,\tilde{\alpha}_2,\tilde{\alpha}_{\mathrm{c}}^{-}}, \label{eq:eigenF10} \\
    \Ket{\psi_{1,1}}&:=\Ket{-\tilde{\alpha}_1,-\tilde{\alpha}_2,\tilde{\alpha}_{\mathrm{c}}^{+}},
    \label{eq:eigenF11}
\end{align}
with
\begin{align}
    \tilde{\alpha}_{\mathrm{c}}^{\pm}&=\frac{\tilde{g}_{1\mathrm{c}}\tilde{\alpha}_1\pm\tilde{g}_{2\mathrm{c}}\tilde{\alpha}_2}{\tilde{\Delta}_{\mathrm{c}}},
\end{align}
are quadruply degenerate eigenstates of $\hat{H}_{0}$ with eigenenergy \TA{$E_{0}=\hbar\sum_{j=1,2}\tilde{K}_j\tilde{\alpha}_j^4/2$}.
These four states are almost orthogonal since the inner product of two coherent states with opposite phases is exponentially small; \TA{$|\braket{\tilde{\alpha}|-\tilde{\alpha}}|=\mathrm{e}^{-2\tilde{\alpha}^2}$}.
We set \TA{$\tilde{\alpha}_j\approx2$} so that \TA{$|\braket{\tilde{\alpha}_j|-\tilde{\alpha}_j}|\approx3\times10^{-4}$} for $j\in\{1,2\}$.
\TA{In order not to apply unwanted $R_X$ gates, we impose $\Braket{\tilde{\alpha}_j|\hat{H}_{X_j}|\tilde{\alpha}_j}=0$, that is,
\begin{linenomath}
\begin{align}
    \tilde{\Delta}_j=\frac{\tilde{g}_{j\mathrm{c}}^2}{\tilde{\Delta}_{\mathrm{c}}}+\frac{17K_j^2}{18\omega_p}\tilde{\alpha}_j^4-\frac{5K_jp_j}{6\omega_p}\tilde{\alpha}_j^2
    \label{eq:noRX2}
\end{align}
\end{linenomath}
for $j\in\{1,2\}$.}
\TA{Since $\omega_p$ is much larger than the other parameters, the terms proportional to $1/\omega_p$ in $\hat{H}_{ZZ}$ can be treated as perturbations to $\hat{H}_{0}$.
By tuning $\tilde{\Delta}_{\mathrm{c}}$, we set $\tilde{g}_{12}-\tilde{g}_{1\mathrm{c}}\tilde{g}_{2\mathrm{c}}/\tilde{\Delta}_{\mathrm{c}}$ so small that the first term in $\hat{H}_{ZZ}$ can  be treated as a perturbation.
We also set $K_{\mathrm{c}}\ll K_j$ and $|\tilde{\alpha}_{\mathrm{c}}^{\pm}|\ll\tilde{\alpha}_j$ so that the second term in $\hat{H}_{ZZ}$ can be treated as a perturbation.
Then, $\hat{H}_{ZZ}$ itself can be treated as a perturbation.}
In the first order of perturbation, the four eigenenergies are calculated as
\begin{align}
    &\quad E^{(1)}_{0,0}/\hbar=E^{(1)}_{1,1}/\hbar \notag \\
    &=E_{0}/\hbar+2\left(\tilde{g}_{12}-\frac{\tilde{g}_{1\mathrm{c}}\tilde{g}_{2\mathrm{c}}}{\tilde{\Delta}_{\mathrm{c}}}\right)\tilde{\alpha}_1\tilde{\alpha}_2
    \notag \\
    &\quad\mbox{}-\frac{\tilde{K}_{\mathrm{c}}}{2}\left(\tilde{\alpha}_{\mathrm{c}}^{+}\right)^4
    -\frac{17K_{\mathrm{c}}^2}{18\omega_p}\left(\tilde{\alpha}_{\mathrm{c}}^{+}\right)^6 \notag \\
    &\quad\mbox{}+\sum_{j=1,2}\frac{2g_{j\mathrm{c}}}{\omega_p}\left[K_j\tilde{\alpha}_j^3\tilde{\alpha}_{\mathrm{c}}^{+}-p_j\tilde{\alpha}_j\tilde{\alpha}_{\mathrm{c}}^{+}+K_{\mathrm{c}}\tilde{\alpha}_j\left(\tilde{\alpha}_{\mathrm{c}}^{+}\right)^3\right] \notag \\
    &\quad\mbox{}-\frac{2g_{12}}{\omega_p}\left[
        K_1\tilde{\alpha}_1^3\tilde{\alpha}_2
        +K_2\tilde{\alpha}_1\tilde{\alpha}_2^3
        -(p_1+p_2)\tilde{\alpha}_1\tilde{\alpha}_2
    \right], \\
    &\quad E^{(1)}_{0,1}/\hbar=E^{(1)}_{1,0}/\hbar \notag \\
    &=E_{0}/\hbar-2\left(\tilde{g}_{12}-\frac{\tilde{g}_{1\mathrm{c}}\tilde{g}_{2\mathrm{c}}}{\tilde{\Delta}_{\mathrm{c}}}\right)\tilde{\alpha}_1\tilde{\alpha}_2
    \notag \\
    &\quad\mbox{}-\frac{\tilde{K}_{\mathrm{c}}}{2}\left(\tilde{\alpha}_{\mathrm{c}}^{-}\right)^4
    -\frac{17K_{\mathrm{c}}^2}{18\omega_p}\left(\tilde{\alpha}_{\mathrm{c}}^{-}\right)^6 \notag \\
    &\quad\mbox{}+\sum_{j=1,2}\frac{2(-1)^{j+1}g_{j\mathrm{c}}}{\omega_p}\left[K_j\tilde{\alpha}_j^3\tilde{\alpha}_{\mathrm{c}}^{-}-p_j\tilde{\alpha}_j\tilde{\alpha}_{\mathrm{c}}^{-}+K_{\mathrm{c}}\tilde{\alpha}_j\left(\tilde{\alpha}_{\mathrm{c}}^{-}\right)^3\right] \notag \\
    &\quad\mbox{}+\frac{2g_{12}}{\omega_p}\left[
        K_1\tilde{\alpha}_1^3\tilde{\alpha}_2
        +K_2\tilde{\alpha}_1\tilde{\alpha}_2^3
        -(p_1+p_2)\tilde{\alpha}_1\tilde{\alpha}_2
    \right],
\end{align}
\TA{where the superscript $(1)$ denotes the first order of perturbation.}
Substituting these four eigenenergies into Eqs.~\TA{\eqref{eq:zetaZZ}}–\eqref{eq:zetaIZ}, we obtain
\begin{align}
    \zeta^{(1)}_{ZZ}&=8\left(\tilde{g}_{12}-\frac{\tilde{g}_{1\mathrm{c}}\tilde{g}_{2\mathrm{c}}}{\tilde{\Delta}_{\mathrm{c}}}\right)\tilde{\alpha}_1\tilde{\alpha}_2
    -\tilde{K}_{\mathrm{c}}\left[
        \left(\tilde{\alpha}_{\mathrm{c}}^{+}\right)^4-\left(\tilde{\alpha}_{\mathrm{c}}^{-}\right)^4
    \right]
    \notag \\
    &\quad\mbox{}-\frac{17K_{\mathrm{c}}^2}{9\omega_p}\left[
        \left(\tilde{\alpha}_{\mathrm{c}}^{+}\right)^6-\left(\tilde{\alpha}_{\mathrm{c}}^{-}\right)^6
    \right]
    \notag \\
    &\quad\mbox{}+\sum_{j=1,2}\frac{4g_{j\mathrm{c}}}{\omega_p}\left[K_j\tilde{\alpha}_j^3\tilde{\alpha}_{\mathrm{c}}^{+}-p_j\tilde{\alpha}_j\tilde{\alpha}_{\mathrm{c}}^{+}+K_{\mathrm{c}}\tilde{\alpha}_j\left(\tilde{\alpha}_{\mathrm{c}}^{+}\right)^3\right]
    \notag \\
    &\quad\mbox{}-\sum_{j=1,2}\frac{4(-1)^{j+1}g_{j\mathrm{c}}}{\omega_p}
    \notag \\
    &\quad\times\left[K_j\tilde{\alpha}_j^3\tilde{\alpha}_{\mathrm{c}}^{-}-p_j\tilde{\alpha}_j\tilde{\alpha}_{\mathrm{c}}^{-}+K_{\mathrm{c}}\tilde{\alpha}_j\left(\tilde{\alpha}_{\mathrm{c}}^{-}\right)^3\right] \notag \\
    &\quad\mbox{}-\frac{8g_{12}}{\omega_p}\left[
        K_1\tilde{\alpha}_1^3\tilde{\alpha}_2
        +K_2\tilde{\alpha}_1\tilde{\alpha}_2^3
        -(p_1+p_2)\tilde{\alpha}_1\tilde{\alpha}_2
    \right]
    \label{eq:zetaZZF1_Kerr}
\end{align}
and $\zeta^{\TA{(1)}}_{ZI}=\zeta^{\TA{(1)}}_{IZ}=0$.
We can cancel $\zeta^{\TA{(1)}}_{ZZ}$ by regulating $\tilde{\varphi}_{\mathrm{c}}^{\rm bias}$ as in Fig.~\ref{fig:zetaZZ}.

\begin{figure}
    \centering
    \includegraphics[width=0.42\textwidth]{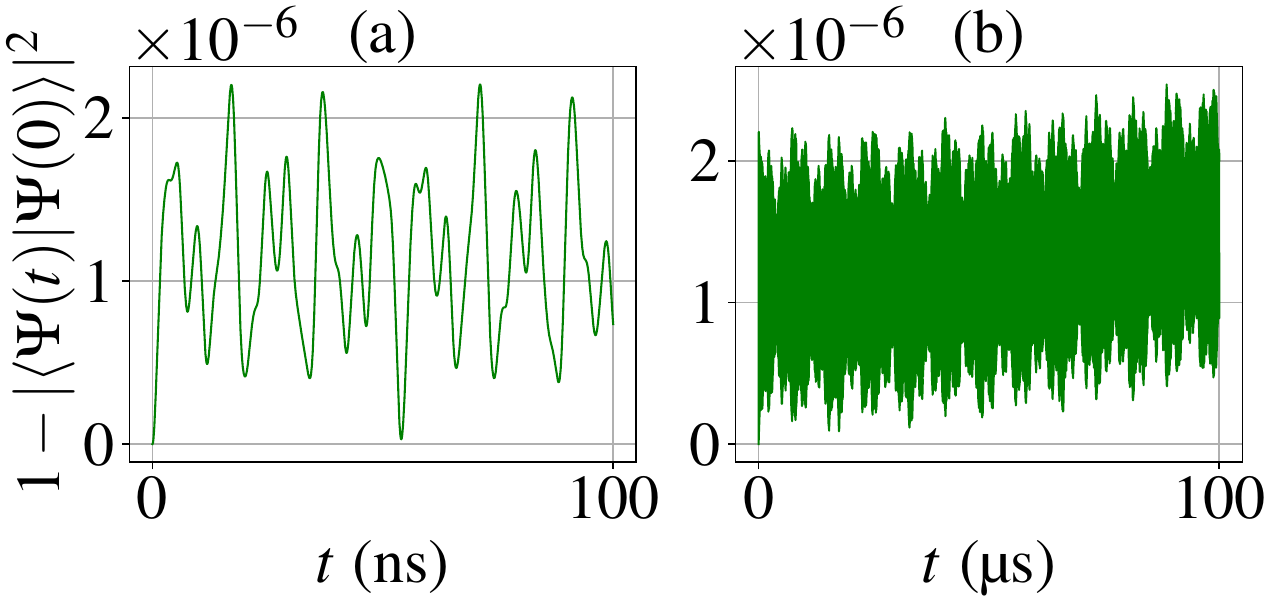}
    \caption{Infidelity $1-|\Braket{\Psi(t)|\Psi(0)}|^2$ when the $ZZ$ coupling should be switched off for (a) $0$\,ns $\leq t\leq100$\,ns and (b) $0$\,µs $\leq t\leq100$\,µs.
    $\Ket{\Psi(t)}=\mathrm{e}^{-\mathrm{i}\hat{H}t/\hbar}\Ket{\Psi(0)}$ with $\hat{H}$ in Eq.~\TA{\eqref{eq:HF2}} and $\Ket{\Psi(0)}$ in Eq.~\eqref{eq:initial1} with $\beta_{0,0}=\beta_{0,1}=\beta_{1,0}=\beta_{1,1}$. 
    The parameters \TA{used} are listed in Table \ref{table:parameter1}.
    \label{fig:residual1}}
\end{figure}

In order to check whether the above scheme for canceling residual $ZZ$ coupling based on the first order of perturbation is effective, we numerically calculate infidelity $1-|\Braket{\Psi(t)|\Psi(0)}|^2$, where $\Ket{\Psi(t)}=\mathrm{e}^{-\mathrm{i}\hat{H}t/\hbar}\Ket{\Psi(0)}$ with $\hat{H}$ in Eq.~\TA{\eqref{eq:HF2}} and $\Ket{\Psi(0)}$ in Eq.~\eqref{eq:initial1} with $\beta_{0,0}=\beta_{0,1}=\beta_{1,0}=\beta_{1,1}$.
For numerical calculations in this Letter, we use Quantum Toolbox in Python (QuTiP).\cite{QuTiP2012,QuTiP2013}
Aiming to satisfy $\zeta^{\TA{(1)}}_{ZZ}\TA{/2\pi}=0$\,\TA{kHz} [Eq.~\TA{\eqref{eq:zetaZZF1_Kerr}}] and Eq.~\TA{\eqref{eq:noRX2}}, we choose the parameter values in Table \ref{table:parameter1}.
We can suppress the residual coupling so that the infidelity oscillates and is less than $\TA{3}\times10^{-\TA{6}}$ \TA{for $0$\,µs $\leq t\leq100$\,µs} as in Fig.~\ref{fig:residual1}, about \TA{three} orders of magnitude smaller than that using two tunable resonators in Ref.~\onlinecite{Aoki2024}.
\begin{table}
    \caption{Parameter values chosen to satisfy $\zeta^{\TA{(1)}}_{ZZ}\TA{/2\pi}=0$\,\TA{kHz} [Eq.~\TA{\eqref{eq:zetaZZF1_Kerr}}] and Eq.~\TA{\eqref{eq:noRX2}}, although there is a slight deviation from Eq.~\TA{\eqref{eq:noRX2}} as \TA{at the bottom}.
    The bold values in the left side are design values, from which the other values in the right side are calculated.
    \label{table:parameter1}}
    \begin{center}
      \begin{tabular}{lclc}
      \hline\hline
      $C_{1}^S=C_2^S$ (fF) & $\boldsymbol{170}$ & $K_1/2\pi$ (MHz) & $13.0$ \\
      $C_{1}^J=C_2^J$ (fF) & $\boldsymbol{30}$ & $K_2/2\pi$ (MHz) & $13.0$ \\
      $N_1=N_2$ & $\boldsymbol{2}$ & $K_{\mathrm{c}}/2\pi$ (MHz) & $1.48$ \\
      $C_{\mathrm{c}}^S$ (fF) & $\boldsymbol{400}$ & $\omega_1/2\pi$ (GHz) & $9.997$ \\
      $C_{\mathrm{c}}^J$ (fF) & $\boldsymbol{30}$ & $\omega_2/2\pi$ (GHz) & $9.996$ \\
      $N_{\mathrm{c}}$ & $\boldsymbol{4}$ & $\omega_{\mathrm{c}}/2\pi$ (GHz) & $10.4$ \\
      $C_{1\mathrm{c}}$ (fF) & $\boldsymbol{2.5}$ & $p_1/2\pi$ (MHz) & $55.1$ \\
      $C_{2\mathrm{c}}$ (fF) & $\boldsymbol{1.3}$ & $p_2/2\pi$ (MHz) & $55.2$ \\
      $C_{12}$ (fF) & $\boldsymbol{0.05}$ & $g_{1\mathrm{c}}/2\pi$ (MHz) & $23.1$ \\
      $E_1^J/h$ (GHz) & $\boldsymbol{341}$ & $g_{2\mathrm{c}}/2\pi$ (MHz) & $12.0$ \\
      $E_2^J/h$ (GHz) & $\boldsymbol{340}$ & $g_{12}/2\pi$ (kHz) & $726$ \\      
      $E_{\mathrm{c}}^J/h$ (THz) & $\boldsymbol{1.14}$ & $\tilde{K}_1/2\pi$ (MHz) & $13.1$ \\
      $\tilde{\varphi}_1^{\rm bias}/2\pi$ & $\boldsymbol{0.250323}$ & $\tilde{K}_2/2\pi$ (MHz) & $13.1$ \\
      $\tilde{\varphi}_2^{\rm bias}/2\pi$ & $\boldsymbol{0.250472}$ & $\tilde{K}_{\mathrm{c}}/2\pi$ (MHz) & $1.48$ \\
      $\tilde{\varphi}_{\mathrm{c}}^{\rm bias}/2\pi$ & $\boldsymbol{2}\times\boldsymbol{10}^{\boldsymbol{-3}}$ & $\tilde{p}_1/2\pi$ (MHz) & $55.2$ \\
      $\epsilon_{p,1}=\epsilon_{p,2}$ & $\boldsymbol{7}\times\boldsymbol{10}^{\boldsymbol{-3}}$ & $\tilde{p}_2/2\pi$ (MHz) & $55.3$ \\
      $\epsilon_{p,\mathrm{c}}$ & $\boldsymbol{0}$ & $\tilde{\Delta}_1/2\pi$ (MHz) & $1.25$ \\
      $\omega_p/2\pi$ (GHz) & $\boldsymbol{19.990741}$ & $\tilde{\Delta}_2/2\pi$ (kHz) & $257$ \\
      $\kappa/2\pi$ (kHz) & $\boldsymbol{4}$ or $\boldsymbol{0}$ & $\tilde{\Delta}_{\mathrm{c}}/2\pi$ (MHz) & $389$ \\
      & & $\tilde{g}_{1\mathrm{c}}/2\pi$ (MHz) & $23.1$ \\
      & & $\tilde{g}_{2\mathrm{c}}/2\pi$ (MHz) & $12.0$ \\
      & & $\tilde{g}_{12}/2\pi$ (kHz) & $711$ \\
      & & $\tilde{\alpha}_{1}$ & $2.06$ \\
      & & $\tilde{\alpha}_{2}$ & $2.05$ \\
      & & $\tilde{\alpha}_{\mathrm{c}}^{+}$ & $0.185$ \\
      & & $\tilde{\alpha}_{\mathrm{c}}^{-}$ & $0.058$ \\
      \multicolumn{3}{c}{$\tilde{\Delta}_1/2\pi-\left[\frac{\tilde{g}_{1\mathrm{c}}^2}{\tilde{\Delta}_{\mathrm{c}}}+\frac{17K_1^2}{18\omega_p}\tilde{\alpha}_1^4-\frac{5K_1p_1}{6\omega_p}\tilde{\alpha}_1^2\right]/2\pi$ (kHz)} & $-1.28$ \\
      \multicolumn{3}{c}{$\tilde{\Delta}_2/2\pi-\left[\frac{\tilde{g}_{2\mathrm{c}}^2}{\tilde{\Delta}_{\mathrm{c}}}+\frac{17K_2^2}{18\omega_p}\tilde{\alpha}_2^4-\frac{5K_2p_2}{6\omega_p}\tilde{\alpha}_2^2\right]/2\pi$ (kHz)} & $-3.77$ \\
      \hline\hline
      \end{tabular}
    \end{center}
\end{table}

\begin{figure}
    \centering
    \includegraphics[width=0.47\textwidth]{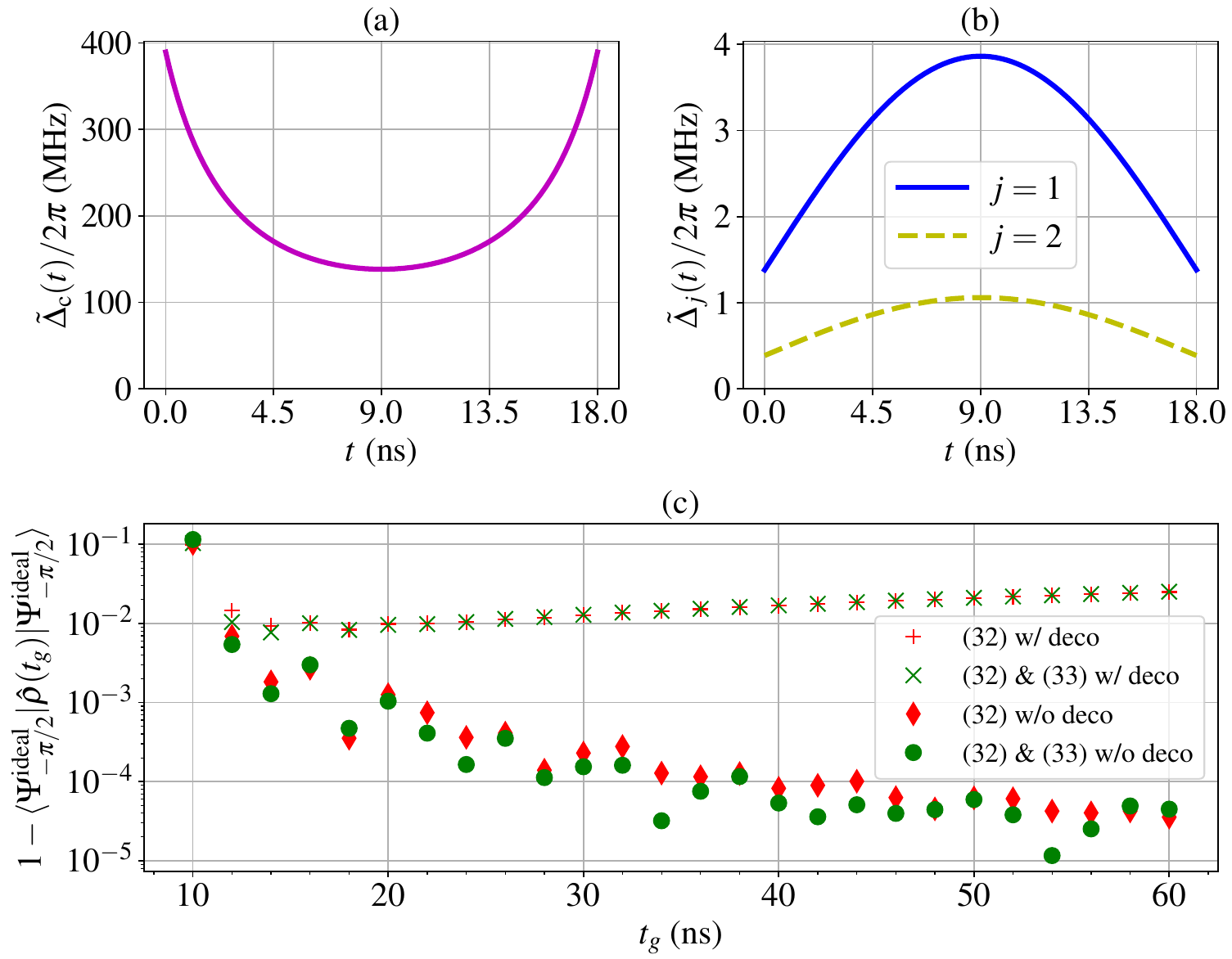}
    \caption{Time variations of (a) $\tilde{\Delta}_{\mathrm{c}}(t)$\TA{,} (b) $\tilde{\Delta}_1(t)$\TA{, and} $\tilde{\Delta}_2(t)$ to satisfy Eqs.~\eqref{eq:nonzero_bsc} and \TA{\eqref{eq:noRX2time}} when an $R_{ZZ}(-\pi/2)$ gate is performed with $t_g=\TA{18}$\,ns.
    (c) \TA{Infidelity} of the $R_{ZZ}(-\pi/2)$ gate \TA{in Eq.~\eqref{eq:gate_infidelity2}}.
    The green circles \TA{(Xs)} are obtained when \TA{all} of $\tilde{\varphi}_{\mathrm{c}}^{\rm bias}(t)$\TA{,} $\tilde{\varphi}_1^{\rm bias}(t)$\TA{, and} $\tilde{\varphi}_2^{\rm bias}(t)$ are tuned to satisfy Eqs.~\eqref{eq:nonzero_bsc} and \TA{\eqref{eq:noRX2time}} \TA{with $\kappa=0$\,Hz ($\kappa/2\pi=4$\,kHz)}. 
    The red diamonds \TA{(plus signs)} are obtained when only $\tilde{\varphi}_{\mathrm{c}}^{\rm bias}(t)$ is tuned to satisfy Eq.~\eqref{eq:nonzero_bsc} \TA{with $\kappa=0$\,Hz ($\kappa/2\pi=4$\,kHz)}; $\tilde{\varphi}_j^{\rm bias}(t)=\tilde{\varphi}_j^{\rm bias}\,\forall t$ for $j\in\{1,2\}$ and Eq.~\TA{\eqref{eq:noRX2time}} is not satisfied.
    The parameters \TA{used} are listed in Table \ref{table:parameter1}.
    \label{fig:RZZ}}
\end{figure}

Next, we partially lift the level degeneracy to switch on the $ZZ$ coupling.
The bias parts $\{\tilde{\varphi}_{\lambda}^{\mathrm{bias}}(t)\}$ are time dependent and so are the related parameters (see the supplementary material).
Though we can apply an $R_{ZZ}(\Theta)$ gate with arbitrary rotation angle $\Theta$, we here evaluate the performance of the $R_{ZZ}(-\pi/2)$ gate.
We tune $\{\tilde{\varphi}_{\lambda}^{\mathrm{bias}}(t)\}$ to satisfy\cite{GotoPRA2016}
\begin{linenomath}
\begin{align}
    \TA{\tilde{g}_{12}-\frac{\tilde{g}_{1\mathrm{c}}\tilde{g}_{2\mathrm{c}}}{\tilde{\Delta}_{\mathrm{c}}(t)}=\tilde{g}_{12}-\frac{\tilde{g}_{1\mathrm{c}}\tilde{g}_{2\mathrm{c}}}{\tilde{\Delta}_{\mathrm{c}}}-\frac{\pi^2}{16\tilde{\alpha}_1\tilde{\alpha}_2t_g}\sin\left(\frac{\pi t}{t_g}\right)}
    \notag \\
    (0\leq t\leq t_g)
    \label{eq:nonzero_bsc}
\end{align}
\end{linenomath}
and the time-dependent version of Eq.~\TA{\eqref{eq:noRX2}},
\begin{linenomath}
    \begin{align}
        \tilde{\Delta}_j(t)=\frac{\tilde{g}_{j\mathrm{c}}^2}{\tilde{\Delta}_{\mathrm{c}}(t)}+\frac{17K_j^2}{18\omega_p}\tilde{\alpha}_j^4-\frac{5K_jp_j}{6\omega_p}\tilde{\alpha}_j^2
        \label{eq:noRX2time}
    \end{align}
\end{linenomath}
with the parameters in Table \ref{table:parameter1}.
For example, when $t_g=\TA{18}$\,ns, \TA{$\{\tilde{\Delta}_{\lambda}(t)\}$} vary as in Fig.~\ref{fig:RZZ}(a,b).
We numerically calculate the infidelity of the $R_{ZZ}(-\pi/2)$ gate,
\TA{
    \begin{linenomath}
\begin{align}
    1-\langle\Psi_{-\pi/2}^{\mathrm{ideal}}|\hat{\rho}(t_g)|\Psi_{-\pi/2}^{\mathrm{ideal}}\rangle,
    \label{eq:gate_infidelity2}
\end{align}
\end{linenomath}
where $\hat{\rho}(t)$ is the density operator of the system and the initial state is $\Ket{\Psi(0)}$ in Eq.~\eqref{eq:initial1} with $\beta_{0,0}=\beta_{0,1}=\beta_{1,0}=\beta_{1,1}$.
We assume a single-photon loss for each subsystem as a major source of decoherence.
Then $\hat{\rho}(t)$ obeys the following Gorini–Kossakowski–Sudarshan–Lindblad (GKSL)-type Markovian master equation: \cite{doi:10.1063/1.522979,lindblad1976generators}
\begin{linenomath}
\begin{align}
    \frac{\mathrm{d}\hat{\rho}(t)}
    {\mathrm{d}t}
    &=-\frac{\mathrm{i}}{\hbar}
    [\hat{H}(t),
    \hat{\rho}(t)]
    \notag \\
    &\mbox{}+\kappa\sum_{\lambda=1,2,\mathrm{c}}\left(
        \hat{a}_{\lambda}\hat{\rho}(t)\hat{a}_{\lambda}^{\dagger}
        -\frac{1}{2}
        \left\{
            \hat{a}_{\lambda}^{\dagger}\hat{a}_{\lambda},\hat{\rho}(t)
        \right\}
    \right),
\end{align}
\end{linenomath}
where $\kappa$ is the single-photon loss rate; $[\bullet,\circ]$ is the commutator; and $\{\bullet,\circ\}$ is the anticommutator.}

\TA{When the decoherence can be neglected ($\kappa=0$\,Hz), the infidelity, which corresponds to green circles in Fig.~\ref{fig:RZZ}(c), tends to decrease accompanied by oscillation with $t_g$.}
We attribute this decrease to the mitigation of unwanted nonadiabatic transitions.
The infidelity is less than \TA{$10^{-3}$} for $t_g=\TA{18}$\,ns.
We also consider a simpler control in which only Eq.~\eqref{eq:nonzero_bsc} is satisfied by tuning only $\tilde{\varphi}_{\mathrm{c}}^{\rm bias}(t)$. 
The infidelity in this case corresponds to red diamonds in Fig.~\ref{fig:RZZ}(c).
Though worse than the green circles \TA{at most of the gate times}, the infidelity is \TA{also} less than \TA{$10^{-3}$} for $t_g=\TA{18}$\,ns.
\TA{When $\kappa/2\pi=4$\,kHz, the infidelities of both controls are almost identical.
The infidelity is approximately $10^{-2}$ for $14$\,ns$\leq t_g\leq24$\,ns.}

In conclusion, we have proposed a $ZZ$-coupling scheme for two Kerr-cat qubits using a tunable resonator.
By engineering the degeneracy of the four relevant states of the two Kerr-cat qubits, we can switch on and off the $ZZ$ coupling.
Our scheme is simpler and achieves smaller residual coupling than the scheme using two tunable resonators in Ref.~\onlinecite{Aoki2024}.
This is because our scheme focuses on energy levels of the whole system, while the scheme in Ref.~\onlinecite{Aoki2024} focuses on those of the effective Hamiltonian of the two resonators;
thus, our scheme enables more precise control of $ZZ$ coupling.
Our scheme can also be applied to superconducting nonlinear asymmetric inductive element (SNAIL)\cite{doi:10.1063/1.4984142}-based Kerr-cat qubits\TA{\cite{Grimm,PhysRevX.14.041049,2024arXiv241104442Q} with modification of the Hamiltonian}.
Integrating our scheme with surface codes for biased noise\cite{PhysRevX.9.041031,Ataides2021} will be an interesting future work.

\section*{SUPPLEMENTARY MATERIAL}
See the supplementary material for \TA{the derivation} of the system Hamiltonian.

\begin{acknowledgments}
The authors are grateful to Hayato Goto for useful discussions.
This Letter is based on results obtained from a project, JPNP16007, commissioned by the New Energy and Industrial Technology Development Organization (NEDO), Japan.
S.M. acknowledges the support from JST [Moonshot R\&D] [Grant Number JPMJMS2061].
\end{acknowledgments}

\section*{AUTHOR DECLARATIONS}
\subsection*{Conflict of Interest}
The authors have no conflicts to disclose.
\subsection*{Author Contributions}
\textbf{Takaaki Aoki:}
Conceptualization (lead);
Data curation (lead);
Formal analysis (lead);
Investigation (lead);
Methodology (lead);
Software (lead);
Validation (lead);
Visualization (lead);
Writing – original draft (lead);
Writing – review \& editing (equal).
\textbf{Akiyoshi Tomonaga:}
Conceptualization (supporting);
Data curation (supporting);
Formal Analysis (supporting);
Investigation (supporting);
Methodology (supporting);
Validation (supporting);
Visualization (supporting);
Writing – review \& editing (equal).
\textbf{Kosuke Mizuno:}
Conceptualization (supporting);
Data curation (supporting);
Formal Analysis (supporting);
Investigation (supporting);
Methodology (supporting);
Validation (supporting);
Visualization (supporting);
Writing – review \& editing (equal).
\textbf{Shumpei Masuda:}
Conceptualization (supporting);
Data curation (supporting);
Formal Analysis (supporting);
Investigation (supporting);
Methodology (supporting);
Project administration (lead);
Supervision (lead);
Validation (supporting);
Visualization (supporting);
Writing – original draft (supporting);
Writing – review \& editing (equal).

\section*{DATA AVAILABILITY}
The data that support the findings of this study are available from the corresponding author upon reasonable request.

\bibliography{1ref}

\widetext
\clearpage
\begin{center}
\textbf{\large Supplemental Material: Residual-\texorpdfstring{$ZZ$}{ZZ}-coupling suppression and fast two-qubit gate for Kerr-cat qubits based on level-degeneracy engineering}
\end{center}
\stepcounter{hoge}
\renewcommand{\theequation}{S\arabic{equation}}

{\large \noindent Derivation of the system Hamiltonian}
\vskip\baselineskip

We derive the Hamiltonian of the circuit in Fig.~2 in the main text in a similar manner to Ref.~\onlinecite{Aoki2024}.
Upward branch flux variables across the left and right Josephson junctions of each SQUID of subsystem $\lambda\in\{1,2,\mathrm{c}\}$ are denoted by $\Phi_{\lambda,L}/N_{\lambda}$ and $\Phi_{\lambda,R}/N_{\lambda}$, respectively;
these variables satisfy $\Phi_{\lambda,L}-\Phi_{\lambda,R}=N_{\lambda}\tilde{\Phi}_{\lambda}(t)$.
Using $\Phi_{\lambda}:=(\Phi_{\lambda,L}+\Phi_{\lambda,R})/2$, we have $\Phi_{\lambda,L}=\Phi_{\lambda}+N_{\lambda}\tilde{\Phi}_{\lambda}(t)/2$ and $\Phi_{\lambda,R}=\Phi_{\lambda}-N_{\lambda}\tilde{\Phi}_{\lambda}(t)/2$.

The kinetic energy of subsystem $\lambda$ is written as
\begin{align}
    T_{\lambda}&=\frac{C_{\lambda}^S}{2}\dot{\Phi}_{\lambda,L}^2+\frac{C_{\lambda}^S}{2}\dot{\Phi}_{\lambda,R}^2
    +N_{\lambda}\left[
        \frac{C_{\lambda}^J}{2}\left(\frac{\dot{\Phi}_{\lambda,L}}{N_{\lambda}}\right)^2
        +\frac{C_{\lambda}^J}{2}\left(\frac{\dot{\Phi}_{\lambda,R}}{N_{\lambda}}\right)^2
    \right]
    =\frac{C_{\lambda}}{2}\dot{\Phi}_{\lambda}^2,
\end{align}
where $C_{\lambda}:=2C_{\lambda}^S+2C_{\lambda}^J/N_{\lambda}$ and we have ignored $N_{\lambda}C_{\lambda}\dot{\tilde{\Phi}}_{\lambda}^2/8$, which does not affect the dynamics of the system.
The kinetic energy of the system is written as
\begin{align}
    T&=\sum_{\lambda=1,2,\mathrm{c}}T_{\lambda}+\frac{C_{12}}{2}(\dot{\Phi}_1-\dot{\Phi}_2)^2
    +\frac{C_{1\mathrm{c}}}{2}(\dot{\Phi}_1-\dot{\Phi}_{\mathrm{c}})^2
    +\frac{C_{2\mathrm{c}}}{2}(\dot{\Phi}_2-\dot{\Phi}_{\mathrm{c}})^2
    =\frac{1}{2}\dot{\boldsymbol{\Phi}}^{\mathrm{T}}M\dot{\boldsymbol{\Phi}}
    =\frac{\phi_0^2}{2}\dot{\boldsymbol{\varphi}}^{\mathrm{T}}M\dot{\boldsymbol{\varphi}},
\end{align}
where
\begin{gather}
    \boldsymbol{\Phi}:=\begin{pmatrix}
        \Phi_{1} \\
        \Phi_{2} \\
        \Phi_{\mathrm{c}}      
    \end{pmatrix},\quad    
    \boldsymbol{\varphi}:=\frac{\boldsymbol{\Phi}}{\phi_0}=\begin{pmatrix}
        \varphi_{1} \\
        \varphi_{2} \\
        \varphi_{\mathrm{c}}      
    \end{pmatrix},
    \quad \varphi_{\lambda}:=\frac{\Phi_{\lambda}}{\phi_0},\\
    M=\begin{pmatrix}
        C_1+C_{12}+C_{1\mathrm{c}} & -C_{12} & -C_{1\mathrm{c}} \\
        -C_{12} & C_2+C_{12}+C_{2\mathrm{c}} & -C_{2\mathrm{c}} \\
        -C_{1\mathrm{c}} & -C_{2\mathrm{c}} & C_{\mathrm{c}}+C_{1\mathrm{c}}+C_{2\mathrm{c}}
    \end{pmatrix}.
\end{gather}
The potential energy of the system is written as
\begin{align}
    U(t)&=-\sum_{\lambda=1,2,\mathrm{c}}N_{\lambda}\left[
        E_{\lambda}^J\cos\left(\frac{\Phi_{\lambda,\mathrm{L}}}{N_{\lambda}\Phi_0}\right)
        +E_{\lambda}^J\cos\left(\frac{\Phi_{\lambda,\mathrm{R}}}{N_{\lambda}\Phi_0}\right)
    \right] \notag \\
    &=-\sum_{\lambda=1,2,\mathrm{c}}2N_{\lambda}E_{\lambda}^J\cos{\left(\frac{\tilde{\varphi}_{\lambda}(t)}{2}\right)}
    \cos{\left(\frac{\varphi_{\lambda}}{N_{\lambda}}\right)},
\end{align}
The Lagrangian of the system in the laboratory frame reads
\begin{align}
    \mathcal{L}(t)=T-U(t).
\end{align}

The conjugate momentum to $\hbar\varphi_{\lambda}$ is denoted by $n_{\lambda}$ and is calculated as
\begin{align}
    \boldsymbol{n}:=\begin{pmatrix}
        n_1 \\ n_2 \\ n_{\mathrm{c}}
    \end{pmatrix}
    =\frac{1}{\hbar}\frac{\partial \mathcal{L}}{\partial\dot{\boldsymbol{\varphi}}}=\frac{\phi_0^2}{\hbar}M\dot{\boldsymbol{\varphi}}.
\end{align}
Legendre transformation gives the classical Hamiltonian in the laboratory frame:
\begin{align}
    H^{\mathrm{lab}}(t)&=\hbar\dot{\boldsymbol{\varphi}}\cdot\boldsymbol{n}-\mathcal{L}(t)=2e^2\boldsymbol{n}^{\mathrm{T}}M^{-1}\boldsymbol{n}+U(t),
    \label{eq:classical_H}
\end{align}
Since $M$ is a symmetric matrix, $M^{-1}$ is also symmetric, and we express $2e^2M^{-1}$ as
\begin{align}
    2e^2M^{-1}=4\begin{pmatrix}
        E_1^C & E_{12}^C & E_{1\mathrm{c}}^C \\
        E_{12}^C & E_2^C & E_{2\mathrm{c}}^C \\
        E_{1\mathrm{c}}^C & E_{2\mathrm{c}}^C & E_{\mathrm{c}}^C 
    \end{pmatrix}.
\end{align}
Quantization $n_{\lambda}\to\hat{n}_{\lambda}$ and $\varphi_{\lambda}\to\hat{\varphi}_{\lambda}$ with $[\hat{\varphi}_{\lambda},\hat{n}_{\lambda'}]=\mathrm{i}\delta_{\lambda,\lambda'}$ leads to the quantum Hamiltonian in the laboratory frame:
\begin{align}
    \hat{H}^{\mathrm{lab}}(t)&=\sum_{\lambda=1,2,\mathrm{c}}\hat{H}^{\mathrm{lab}}_{\lambda}(t)+\hat{H}_{\mathrm{I}}^{\mathrm{lab}},
    \label{eq:Hamiltonian_total1} \\
    \hat{H}^{\mathrm{lab}}_{\lambda}(t)&=4E_{\lambda}^C\hat{n}_{\lambda}^2-2N_{\lambda}E_{\lambda}^J\cos{\left(\frac{\tilde{\varphi}_{\lambda}(t)}{2}\right)}
    \cos{\left(\frac{\hat{\varphi}_{\lambda}}{N_{\lambda}}\right)},
    \label{eq:Hamiltonian_lambda1} \\
    \hat{H}_{\mathrm{I}}^{\mathrm{lab}}&=\sum_{j=1,2}8E_{j\mathrm{c}}^C\hat{n}_j\hat{n}_{\mathrm{c}}+8E_{12}^C\hat{n}_1\hat{n}_2.
    \label{eq:Hamiltonian_I1}
\end{align}
We set parameters so that $E_{\lambda}^C\ll N_{\lambda}E_{\lambda}^J\cos{(\tilde{\varphi}_{\lambda}(t)/2)}\,\forall t$, which justifies expanding $\cos(\hat{\varphi}_{\lambda}/N_{\lambda})$ to the fourth order in $\hat{\varphi}_{\lambda}/N_{\lambda}$:
\begin{align}
    \cos{\left(\frac{\hat{\varphi}_{\lambda}}{N_{\lambda}}\right)}\approx1-\frac{1}{2N_{\lambda}^2}\hat{\varphi}_{\lambda}^2+\frac{1}{24N_{\lambda}^4}\hat{\varphi}_{\lambda}^4.
    \label{eq:cos_fourth}
\end{align}
Since $|\epsilon_{p,\lambda}|\ll1$, we have
\begin{align}
    \cos{\left(\frac{\tilde{\varphi}_{\lambda}(t)}{2}\right)}&\approx\cos{\left(\frac{\tilde{\varphi}_{\lambda}^{\mathrm{bias}}(t)}{2}\right)}+\pi\epsilon_{p,\lambda}\sin{\left(\frac{\tilde{\varphi}_{\lambda}^{\mathrm{bias}}(t)}{2}\right)}\cos{(\omega_pt)}.
\end{align}
Defining
\begin{align}
    \tilde{E}_{\lambda}^{J,\mathrm{bias}}(t)&:=2E_{\lambda}^J\cos{\left(\frac{\tilde{\varphi}_{\lambda}^{\mathrm{bias}}(t)}{2}\right)},
    \label{eq:EJ_tilde_bias1} \\
    \tilde{E}_{\lambda}^{J,\mathrm{pump}}(t)&:=2\pi\epsilon_{p,\lambda}E_{\lambda}^J\sin{\left(\frac{\tilde{\varphi}_{\lambda}^{\mathrm{bias}}(t)}{2}\right)}\cos{(\omega_pt)},
    \label{eq:EJ_tilde_pump1}
\end{align}
we obtain
\begin{align}
    \hat{H}_{\lambda}^{\mathrm{lab}}(t)&\approx4E_{\lambda}^C\hat{n}_{\lambda}^2
    +\frac{\tilde{E}_{\lambda}^{J,\mathrm{bias}}(t)}{2N_{\lambda}}\hat{\varphi}_{\lambda}^2
    +\frac{\tilde{E}_{\lambda}^{J,\mathrm{pump}}(t)}{2N_{\lambda}}\hat{\varphi}_{\lambda}^2
    -\frac{\tilde{E}_{\lambda}^{J,\mathrm{bias}}(t)}{24N_{\lambda}^3}\hat{\varphi}_{\lambda}^4,
    \label{eq:Hamiltonian_lambda2}
\end{align}
where we have ignored the smallest quantum-number term, $\tilde{E}_{\lambda}^{J,\mathrm{pump}}(t)\hat{\varphi}_{\lambda}^4/(24N_{\lambda}^3)$, and classical-number terms.

Inserting
\begin{align}
    \hat{\varphi}_{\lambda}&=
    \left(\frac{2N_{\lambda}E_{\lambda}^C}{\tilde{E}_{\lambda}^{J,\mathrm{bias}}(0)}
    \right)^{1/4}
    \left(\hat{a}_{\lambda}+\hat{a}_{\lambda}^{\dagger}\right),
    \label{eq:hat_phi1}\\
    \hat{n}_{\lambda}&=
    -\frac{\mathrm{i}}{2}
    \left(\frac{\tilde{E}_{\lambda}^{J,\mathrm{bias}}(0)}{2N_{\lambda}E_{\lambda}^C}
    \right)^{1/4}
    \left(\hat{a}_{\lambda}-\hat{a}_{\lambda}^{\dagger}\right),
    \label{eq:hat_n1}
\end{align}
where $\hat{a}_{\lambda}$ and $\hat{a}_{\lambda}^{\dagger}$ are the annihilation and creation operators, into Eq.~\eqref{eq:Hamiltonian_lambda2} gives
\begin{align}
    \hat{H}_{\lambda}^{\mathrm{lab}}(t)/\hbar&\approx
    \omega_{\lambda}^{(0)}(t)\hat{a}_{\lambda}^{\dag}\hat{a}_{\lambda}
    +\frac{\omega_{\lambda}^{(0)}(t)-\omega_{\lambda}^{(0)}(0)}{2}\left(\hat{a}_{\lambda}^{\dag2}+\hat{a}_{\lambda}^{2}\right)
    -\frac{K_{\lambda}(t)}{12}\left(\hat{a}_{\lambda}+\hat{a}_{\lambda}^{\dagger}\right)^4
    \notag \\
    &\quad\mbox{}+p_{\lambda}(t)\cos{(\omega_pt)}\left(\hat{a}_{\lambda}+\hat{a}_{\lambda}^{\dagger}\right)^2,
\end{align}
where
\begin{align}
    \hbar\omega_{\lambda}^{(0)}(t)&:=\sqrt{\frac{2E_{\lambda}^C\tilde{E}_{\lambda}^{J,\mathrm{bias}}(0)}{N_{\lambda}}}
    \left(\frac{\tilde{E}_{\lambda}^{J,\mathrm{bias}}(t)}{\tilde{E}_{\lambda}^{J,\mathrm{bias}}(0)}+1\right),
    \label{eq:omega_c_lambda_0_t1} \\
    \hbar K_{\lambda}(t)&:=\frac{E_{\lambda}^C\tilde{E}_{\lambda}^{J,\mathrm{bias}}(t)}{N_{\lambda}^2\tilde{E}_J^{\lambda,\mathrm{bias}}(0)},
    \label{eq:K_lambda_t1} \\
    \hbar p_{\lambda}(t)&:=\pi\epsilon_{p,\lambda}E_{\lambda}^J\sqrt{\frac{2E_{\lambda}^C}{N_{\lambda}\tilde{E}_{\lambda}^{J,\mathrm{bias}}(0)}}\sin{\left(\frac{\tilde{\varphi}_{\lambda}^{\mathrm{bias}}(t)}{2}\right)},
    \label{eq:p_lambda_t1}
\end{align}
and we have omitted a classical-number term.
Since the resonator does not have the pump part of the dimensionless magnetic flux, $\epsilon_{p,\mathrm{c}}=0$, we have $p_{\mathrm{c}}(t)=0\;\forall t$.
Substituting Eq.~\eqref{eq:hat_n1} into Eq.~\eqref{eq:Hamiltonian_I1} leads to
\begin{align}
    \hat{H}_{\mathrm{I}}^{\mathrm{lab}}/\hbar&=-\sum_{j=1}^2g_{j\mathrm{c}}
    \left(\hat{a}_{j}-\hat{a}_{j}^{\dagger}\right)
    \left(\hat{a}_{\mathrm{c}}-\hat{a}_{\mathrm{c}}^{\dagger}\right)
    -g_{12}\left(\hat{a}_{1}-\hat{a}_{1}^{\dagger}\right)
    \left(\hat{a}_{2}-\hat{a}_{2}^{\dagger}\right)
\end{align}
where
\begin{align}
    \hbar g_{j\mathrm{c}}=\sqrt{2}E_{j\mathrm{c}}^C\left(\frac{\tilde{E}_{j}^{J,\mathrm{bias}}(0)\tilde{E}_{\mathrm{c}}^{J,\mathrm{bias}}(0)}{N_{j}N_{\mathrm{c}}E_{j}^CE_{\mathrm{c}}^C}
    \right)^{1/4}, \quad
    \hbar g_{12}=\sqrt{2}E_{12}^C\left(\frac{\tilde{E}_{1}^{J,\mathrm{bias}}(0)\tilde{E}_{2}^{J,\mathrm{bias}}(0)}{N_{1}N_{2}E_{1}^CE_{2}^C}
    \right)^{1/4}.
\end{align}

Moving on to a rotating frame at frequency $\omega_p/2$, the Hamiltonian of the system can be written as
\begin{align}
    \hat{H}^{\mathrm{R}}(t)&=\hat{R}^{\dagger}(t)\hat{H}^{\mathrm{lab}}(t)\hat{R}(t)-\mathrm{i}\hbar\hat{R}^{\dagger}(t)\frac{\mathrm{d}\hat{R}(t)}{\mathrm{d}t}
    =\sum_{m=0,\pm1,\pm2}\hat{H}^{\mathrm{R}}_m(t)\mathrm{e}^{\mathrm{i}m\omega_pt},
\end{align}
where
\begin{align}
    \hat{R}(t)&=\exp\left[-
        \mathrm{i}\frac{\omega_p}{2}t\sum_{\lambda=1,2,\mathrm{c}}\hat{a}_{\lambda}^{\dag}\hat{a}_{\lambda}
    \right], \\
    \hat{H}^{\mathrm{R}}_0(t)/\hbar&\mbox{}=\sum_{j=1,2}\left[-\frac{K_j(t)}{2}\hat{a}_j^{\dag2}\hat{a}_j^2+\frac{p_j(t)}{2}(\hat{a}_j^{\dag2}+\hat{a}_j^2)+\Delta_j(t)\hat{a}_j^{\dag}\hat{a}_j\right] \notag \\
    &\quad\mbox{}-\frac{K_{\mathrm{c}}(t)}{2}\hat{a}_{\mathrm{c}}^{\dag2}\hat{a}_{\mathrm{c}}^2+\Delta_{\mathrm{c}}(t)\hat{a}_{\mathrm{c}}^{\dag}\hat{a}_{\mathrm{c}} \notag \\
    &\quad\mbox{}+\sum_{j=1,2}g_{j\mathrm{c}}(\hat{a}_j^{\dag}\hat{a}_{\mathrm{c}}+\hat{a}_j\hat{a}_{\mathrm{c}}^{\dag})
    +g_{12}(\hat{a}_1^{\dag}\hat{a}_2+\hat{a}_1\hat{a}_2^{\dag}), \label{eq:H0Rt}\\
    \hat{H}^{\mathrm{R}}_1(t)/\hbar&=\sum_{j=1,2}\left[p_j(t)\hat{a}_j^{\dag}\hat{a}_j-\frac{K_j(t)}{3}\hat{a}_j^{\dag3}\hat{a}_j+\frac{\omega_j^{(0)}(t)-\omega_j^{(0)}(0)-K_j(t)}{2}\hat{a}_j^{\dag2}\right] \notag \\
    &\quad\mbox{}-\frac{K_{\mathrm{c}}(t)}{3}\hat{a}_{\mathrm{c}}^{\dag3}\hat{a}_{\mathrm{c}}+\frac{\omega_{\mathrm{c}}^{(0)}(t)-\omega_{\mathrm{c}}^{(0)}(0)-K_{\mathrm{c}}(t)}{2}\hat{a}_{\mathrm{c}}^{\dag2} \notag \\
    &\quad\mbox{}-\sum_{j=1,2}g_{j\mathrm{c}}\hat{a}_j^{\dag}\hat{a}_{\mathrm{c}}^{\dag}-g_{12}\hat{a}_1^{\dag}\hat{a}_2^{\dag}, \\
    \hat{H}^{\mathrm{R}}_{-1}(t)/\hbar&=\sum_{j=1,2}\left[p_j(t)\hat{a}_j^{\dag}\hat{a}_j-\frac{K_j(t)}{3}\hat{a}_j^{\dag}\hat{a}_j^3+\frac{\omega_j^{(0)}(t)-\omega_j^{(0)}(0)-K_j(t)}{2}\hat{a}_j^{2}\right] \notag \\
    &\quad\mbox{}-\frac{K_{\mathrm{c}}(t)}{3}\hat{a}_{\mathrm{c}}^{\dag}\hat{a}_{\mathrm{c}}^3+\frac{\omega_{\mathrm{c}}^{(0)}(t)-\omega_{\mathrm{c}}^{(0)}(0)-K_{\mathrm{c}}(t)}{2}\hat{a}_{\mathrm{c}}^{2} \notag \\
    &\quad\mbox{}-\sum_{j=1,2}g_{j\mathrm{c}}\hat{a}_j\hat{a}_{\mathrm{c}}-g_{12}\hat{a}_1\hat{a}_2, \\
    \hat{H}^{\mathrm{R}}_{2}(t)/\hbar&=\sum_{j=1,2}\left[\frac{p_j(t)}{2}\hat{a}_j^{\dag2}-\frac{K_j(t)}{12}\hat{a}_j^{\dag4}\right]-\frac{K_{\mathrm{c}}(t)}{12}\hat{a}_{\mathrm{c}}^{\dag4}, \\
    \hat{H}^{\mathrm{R}}_{-2}(t)/\hbar&=\sum_{j=1,2}\left[\frac{p_j(t)}{2}\hat{a}_j^{2}-\frac{K_j(t)}{12}\hat{a}_j^{4}\right]-\frac{K_{\mathrm{c}}(t)}{12}\hat{a}_{\mathrm{c}}^{4}, \label{eq:H2Rt}
\end{align}
with $\Delta_{\lambda}(t):=\omega_{\lambda}(t)-\omega_p/2$ and $\omega_{\lambda}(t):=\omega_{\lambda}^{(0)}(t)-K_{\lambda}(t)$.
Note that $\hat{H}^{\mathrm{R}}_0(t)$ is the Hamiltonian of the system under the rotating-wave approximation.
At first we assume that the bias parts of the dimensionless magnetic fluxes are time independent so that $\{\hat{H}^{\mathrm{R}}_m|m\in\{0,\pm1,\pm2\}\}$ are time independent.
Then, $\hat{H}^{\mathrm{R}}(t)$ is time-periodic with period $T=2\pi/\omega_p$, and the time-evolution operator takes the form\cite{PhysRev.138.B979,PhysRevA.68.013820,Eckardt_2015,Bukov04032015,RevModPhys.89.011004,GarciaMata2024effectiveversus}
\begin{align}
    \hat{U}(t)&=\mathcal{T}\exp\left[-\frac{\mathrm{i}}{\hbar}\int_0^t\hat{H}^{\mathrm{R}}(s)\mathrm{d}s\right]
    =\hat{U}_F(t)\exp\left[-\frac{\mathrm{i}}{\hbar}\hat{H}_Ft\right]\hat{U}_F^{\dag}(0),
\end{align}
where $\mathcal{T}$ is the time-ordering operator; $\hat{U}_F(t)=\mathrm{e}^{-\mathrm{i}\hat{K}(t)}$ is a micromotion operator, describes a time-periodic component of the dynamics, and satisfies $\hat{U}_F(t+T)=\hat{U}_F(t)$; $\hat{K}(t)$ is a Hermitian time-periodic kick operator;\cite{PhysRevX.4.031027} and
\begin{align}
    \hat{H}_F=\hat{U}_F^{\dagger}(t)\hat{H}^{\mathrm{R}}(t)\hat{U}_F(t)-\mathrm{i}\hbar\hat{U}_F^{\dagger}(t)\frac{\mathrm{d}\hat{U}_F(t)}{\mathrm{d}t}
\end{align}
is a time-independent effective Floquet Hamiltonian.

We set $\omega_p$ much larger than the other parameters in $\hat{H}^{\mathrm{R}}(t)$, and approximate $\hat{H}_F$ and $\hat{K}(t)$ using the Van Vleck (high-frequency) expansion in powers of $1/\omega_p$ to the first order:\cite{PhysRevA.68.013820,Eckardt_2015,PhysRevX.4.031027,Bukov04032015,RevModPhys.89.011004,PhysRevLett.129.100601}
\begin{align}
    \hat{H}_F&\approx\hat{H}_F^{(0)}+\hat{H}_F^{(1)}=:\hat{H}, \label{eq:HF} \\
    \hat{H}_F^{(0)}&=\hat{H}^{\mathrm{R}}_0, \label{eq:HF0th} \\
    \hat{H}_F^{(1)}&=\sum_{m=\pm1,\pm2}\frac{\hat{H}^{\mathrm{R}}_m\hat{H}^{\mathrm{R}}_{-m}}{m\hbar\omega_p}, \label{eq:HF1st} \\
    \hat{K}(t)&\approx\hat{K}^{(0)}(t)+\hat{K}^{(1)}(t), \\
    \hat{K}^{(0)}(t)&=0, \\
    \hat{K}^{(1)}(t)&=-\sum_{m=\pm1,\pm2}\frac{\hat{H}^{\mathrm{R}}_m\mathrm{e}^{\mathrm{i}m\omega_pt}}{m\hbar\omega_p}.
\end{align}
Combining Eqs.~\eqref{eq:H0Rt}–\eqref{eq:H2Rt} and Eqs.~\eqref{eq:HF}–\eqref{eq:HF1st}, we obtain
\begin{align}
    \hat{H}&=\sum_{j=1,2}\hat{H}_{j}+\hat{H}_{\mathrm{c}}+\hat{H}_{\mathrm{I}}, \\
    \hat{H}_{j}/\hbar&=-\frac{\tilde{K}_j}{2}\hat{a}_j^{\dag2}\hat{a}_j^2+\frac{\tilde{p}_j}{2}(\hat{a}_j^{\dag2}+\hat{a}_j^2)+\tilde{\Delta}_j\hat{a}_j^{\dag}\hat{a}_j-\frac{17K_j^2}{18\omega_p}\hat{a}_j^{\dag3}\hat{a}_j^3+\frac{5K_jp_j}{6\omega_p}(\hat{a}_j^{\dag3}\hat{a}_j+\hat{a}_j^{\dag}\hat{a}_j^3), \\
    \hat{H}_{\mathrm{c}}/\hbar&=-\frac{\tilde{K}_{\mathrm{c}}}{2}\hat{a}_{\mathrm{c}}^{\dag2}\hat{a}_{\mathrm{c}}^2+\tilde{\Delta}_{\mathrm{c}}\hat{a}_{\mathrm{c}}^{\dag}\hat{a}_{\mathrm{c}}-\frac{17K_{\mathrm{c}}^2}{18\omega_p}\hat{a}_{\mathrm{c}}^{\dag3}\hat{a}_{\mathrm{c}}^3, \\
    \hat{H}_{\mathrm{I}}/\hbar&=\sum_{j=1,2}\tilde{g}_{j\mathrm{c}}(\hat{a}_j^{\dag}\hat{a}_{\mathrm{c}}+\hat{a}_j\hat{a}_{\mathrm{c}}^{\dag})
    +\tilde{g}_{12}(\hat{a}_1^{\dag}\hat{a}_2+\hat{a}_1\hat{a}_2^{\dag}) \notag \\
    &\quad\mbox{}-\sum_{j=1,2}\frac{g_{j\mathrm{c}}}{\omega_p}[K_j(\hat{a}_j^{\dag2}\hat{a}_j\hat{a}_{\mathrm{c}}+\hat{a}_j^{\dag}\hat{a}_j^2\hat{a}_{\mathrm{c}}^{\dag})-p_j(\hat{a}_j^{\dag}\hat{a}_{\mathrm{c}}^{\dag}+\hat{a}_j\hat{a}_{\mathrm{c}})+K_{\mathrm{c}}(\hat{a}_j^{\dag}\hat{a}_{\mathrm{c}}^{\dag}\hat{a}_{\mathrm{c}}^{2}+\hat{a}_j\hat{a}_{\mathrm{c}}^{\dag2}\hat{a}_{\mathrm{c}})] \notag \\
    &\quad\mbox{}-\frac{g_{12}}{\omega_p}[K_1(\hat{a}_1^{\dag2}\hat{a}_1\hat{a}_2+\hat{a}_1^{\dag}\hat{a}_1^2\hat{a}_2^{\dag})+K_2(\hat{a}_1^{\dag}\hat{a}_2^{\dag}\hat{a}_2^2+\hat{a}_1\hat{a}_2^{\dag2}\hat{a}_2)-(p_1+p_2)(\hat{a}_1^{\dag}\hat{a}_2^{\dag}+\hat{a}_1\hat{a}_2)],
\end{align}
where
\begin{align}
    \tilde{K}_j&=\left(1+\frac{17K_j}{2\omega_p}\right)K_j, \\
    \tilde{p}_j&=\left(1+\frac{5K_j}{2\omega_p}\right)p_j, \\
    \tilde{\Delta}_j&=\Delta_j-\frac{1}{\omega_p}\left(4K_j^2+\frac{p_j^2}{2}-g_{j\mathrm{c}}^2-g_{12}^2\right), \\
    \tilde{K}_{\mathrm{c}}&=\left(1+\frac{17K_{\mathrm{c}}}{2\omega_p}\right)K_{\mathrm{c}}, \\
    \tilde{\Delta}_{\mathrm{c}}&=\Delta_{\mathrm{c}}-\frac{1}{\omega_p}\left(4K_{\mathrm{c}}^2-g_{1\mathrm{c}}^2-g_{2\mathrm{c}}^2\right), \\
    \tilde{g}_{j\mathrm{c}}&=g_{j\mathrm{c}}-\frac{1}{\omega_p}\left[\frac{g_{1\mathrm{c}}g_{2\mathrm{c}}g_{12}}{g_{j\mathrm{c}}}+g_{j\mathrm{c}}(K_j+K_{\mathrm{c}})\right], \\
    \tilde{g}_{12}&=g_{12}-\frac{1}{\omega_p}\left[g_{1\mathrm{c}}g_{2\mathrm{c}}+g_{12}(K_1+K_2)\right].
\end{align}

\end{document}